\documentstyle[aps,twocolumn,epsfig,rotating]{revtex}

\begin{document}




\begin{title} {\bf Superfluid density of high-$T_{c}$ cuprate 
systems: implication on condensation mechanisms, heterogeneity and phase
diagram\/}
\end{title}

\author{Y.J.~Uemura
\footnote{
E-mail: tomo@lorentz.phys.columbia.edu}
}

\address{Department of Physics, Columbia University, 
New York, NY 10027, USA}
\vspace{-5truemm}


\maketitle

\vfill\eject

\newpage

\begin{abstract}
Extensive muon spin relaxation ($\mu$SR)
measurements have been performed to determine 
the magnetic field penetration depth $\lambda$
in high $T_{c}$ cuprate superconductors with simple hole doping,
Zn-doping, overdoping, and formation of static SDW nano islands.
System dependence of $n_{s}/m^{*}$ (superconducting
carrier density / effective mass) reveals universal correlations
between $T_{c}$ and $n_{s}/m^{*}$ in all these cases with/without
perturbation.  
Evidence for spontaneous and microscopic phase
separation into normal and superconducting regions was 
obtained in the cases with strong perturbation, i.e.,   
Zn-doping (swiss cheese model), overdoping, and coexisting
magnetic and superconducting states (SDW nano islands).
The length scale of this heterogeneity is shown to be comparable to 
the in-plane coherence length.
We discuss implication of these results on 
condensation mechanisms of HTSC systems, resorting to an   
analogy with pure $^{4}$He and $^{4}$He/$^{3}$He mixture   
films on regular and porous media, reminding essential 
features of Bose-Einstein, 
BCS and Kosterlitz-Thouless condensation/transition
in 3-d and 2-d systems, and
comparing models of BE-BCS crossover and phase fluctuations. 
Combining the $\mu$SR results on 
$n_{s}/m^{*}$ and the pseudo-gap behavior, 
we propose a new phase diagram for HTSC,
characterized by:
(1) the $T^{*}$ line that represents pair formation;
(2) disappearance of this line above the critical hole 
concentration $x=x_{c}$;
(3) in the underdoped region between $T_{c}$ and $T^{*}$,
there exists another line $T_{dyn}$ which corresponds to the onset
of dynamic superconductivity with superconducting phase fluctuations;
and 
(4) the overdoped region being phase separated between 
hole-poor superfluid and
hole-rich normal fermion metal regions.
Finally, we elucidate anomalous reduction of superfluid spectral
weight in the crossover from superconducting to metallic
ground states found not only in overdoped HTSC cuprates but also in 
pressurized organic BEDT and A$_{3}$C$_{60}$ fulleride superconductors.
\end{abstract}


%



\narrowtext
\section{Introduction}
Muon spin relaxation ($\mu$SR) technique [1-3] has made significant 
contributions to studies of high-$T_{c}$ superconductors (HTSC).  
Measurements in transverse-fields (TF-$\mu$SR) provide 
a reliable way to determine the magnetic field
penetration depth $\lambda$ of type-II superconductors
and to infer details of the flux vortex lattice, while
measurements in zero-filed (ZF-$\mu$SR) allow studies of static
magnetism, leading to determination of the 
phase diagrams, local spin structures,
and volume fraction of static magnetism.  
In this article, we provide a perspective view on
development of these measurements in HTSC systems and their
implications on condensation mechanisms, heterogeneity and phase 
diagrams, focusing on the behavior of 
$n_{s}/m^{*}$ (superconducting carrier density / effective mass).
Since the parameter $n_{s}/m^{*}$ in HTSC systems plays a role similar 
to the superfluid density in superfluid He systems, 
we refer to it as ``superfluid density'' in the title 
and the text of this paper.
   
In the following section, Section II, 
we start with the results of the penetration depth 
measurements of hole-doped cuprate systems, which exhibit  
universal correlations between $T_{c}$ and $n_{s}/m^{*}$.  
We discuss these correlations in terms of energy
scales of superconducting carriers.
Then, we look into the cases involving heterogeneity, i.e., 
(Cu,Zn) substitution and overdoping in Section III, and 
systems with coexisting superconductivity and 
static magnetism, with the formation of magnetic nano-islands, 
in section IV.  
In all these cases, the superfluid density has a
trade-off with the volume of non-superconducting regions created by
perturbation and/or phase separation. We shall see that the
universal correlations between $T_{c}$ and the superfluid density
$n_{s}/m^{*}$ are followed not
only by the simple hole doped cuprates but also by these HTSC systems
with microscopic phase separation.   

Thin films of superfluid $^{4}$He and $^{4}$He/$^{3}$He mixtures 
represent another set of systems where the superfluid transition
temperature $T_{c}$ is strongly correlated with the superfluid
density.  In section V, we compare the results of HTSC systems
with superfluid He films in normal and porous media.
 
Based on the correlations between $T_{c}$ and $n_{s}/m^{*}$
and the pseudogap phenomena in the underdoped region,
some models/conjectures have been presented in terms of the
condensation mechanisms of HTSC systems. 
To elucidate these models, in section VI, 
we will first consider differences and similarities among
Bose-Einstein (BE) condensation, BCS condensation and 
Kosterlitz-Thouless (KT) transition.  Then, we will  
compare models of BE-BCS crossover and 
phase fluctuations, taking into account relevance to 
the KT transition and distinction between
pair-formation and dynamic superconductivity
in the pseudo-gap regime.  We will also 
introduce a phase diagram involving phase separation in the overdoped region
in section VII, followed by a summary in section VIII.

\section{Correlations between $T_{c}$ and $n_{s}/m^{*}$ in hole-doped HTSC}

In type-II superconductors, the external magnetic field, between
$H_{c1}$ and $H_{c2}$, forms a lattice of 
flux vortices, resulting in inhomogeneous internal field distributions. 
The width of this distribution is proportional to 
the muon spin relaxation rate $\sigma$ in TF-$\mu$SR, which is related to the
penetration depth $\lambda$ and $n_{s}/m^{*}$ as 
$$\sigma \propto \lambda^{-2} = [4\pi n_{s}e^{2}/c^{2}] \times 
[1/(1+\xi/{\it l\/})]\eqno{(1)},$$
as given by the London equation.
In superconducting systems in the clean limit, where the coherence 
length $\xi$ is much shorter than the mean free path {\it l\/}, the 
relaxation rate $\sigma$ is proportional to $n_{s}/m^{*}$. 
The inhomogeneity of the internal field is due to partial 
screening of the external field by the supercurrent.
So, it is quite natural that the field width and $\sigma$ are
proportional to the ``supercurrent density'' $n_{s}/m^{*}$, in the 
same sense as the conductivity of normal metals is proportional 
to $n_{n}/m^{*}$, where $n_{n}$ denotes the normal state carrier density.

Generally, TF-$\mu$SR studies in type-II superconductors provide 
information on:
(a) the temperature dependence of $\lambda$; (b) absolute values of 
$\lambda$ at $T \rightarrow 0$; (c) the coherence length $\xi$ via analyses of
local field distribution; and (d) vortex lattice properties.
Among these, use of single crystal specimens is essential in 
(a), (c) and (d).  For example, early $\mu$SR results of $\lambda$ on 
ceramic samples of HTSC systems mostly exhibited behavior consistent 
with isotropic energy gap, and it is only after studies of high-quality 
single crystal specimens that the d-wave nature was confirmed in 
temperature dependence of $\lambda$.  For details of (a), (c) and (d), 
readers are referred to refs. [3-4].  

In contrast, the results 
on ceramic specimens have been very useful in comparing absolute 
values of $n_{s}/m^{*}$ in different systems. 
Ceramic specimens provide advantages over single crystal specimens in
homogeneity of doped hole concentrations and in
being less affected by crossover from the 3-d to 2-d vortex lattice. 
In this article, we shall focus on the system and
doping dependence of the superfluid density, i.e., (b).  
In HTSC systems with a large
anisotropy, the relaxation rate $\sigma$ measured in ceramic
specimens is determined predominantly by the supercurrent 
flowing in the conducting ab-planes [5]. Thus, the  
results of $\sigma$ from ceramic specimens 
should be regarded as reflecting the in-plane penetration 
depth $\lambda_{ab}$ 
and the in-plane effective mass $m^{*}_{ab}$.

Figure 1 shows a plot of $\sigma(T\rightarrow 0)$ versus $T_{c}$
of various different HTSC systems [6-11].  In this figure, 
results from simple hole-doped HTSC systems [6,7] are shown with 
open symbols. 
With increasing hole concentration, $T_{c}$ increases linearly with
$n_{s}/m^{*}$ in the underdoped region, and then shows a saturation.
The slope of this linear relationship in the underdoped region 
is common to various different 
series of HTSC systems.  
These results have also been 
confirmed in several other $\mu$SR measurements 
[12-16].
This observation alone already implies that
$n_{s}/m^{*}$ can be an important determining factor for $T_{c}$ in 
HTSC systems.    We also find in Fig. 1 that the ratios of 
$T_{c}$ versus $n_{s}/m^{*}$ for some other superconductors [17,18]
are not quite far from the value for HTSC systems.  

If an independent estimate for the effective mass $m^{*}$ is
available, the $\mu$SR measurements give the superconducting
carrier density $n_{s}$.  Then, one can calculate the number of
carries existing in a region of coherence length squared on the 
conducting plane.  For typical BCS superconductors, such as Sn or Pb,
one finds more than 10,000 pairs overlapping with one another in the
$\pi\xi^{2}$ area.  For superfluid He, which is the well-known case for
Bose-Einstein condensation, we find that each boson exist without
overlapping with each other: i.e., one pair per $\pi\xi^{2}$.
The systems following the linear relationship in Fig. 1 have
3 to 6 pairs overlapping within the $\pi\xi^{2}$ area,
as illustrated in Fig. 2.  
This feature encourages us to consider HTSC and some
novel superconductors   
in a crossover from BE to BCS condensation.

By knowing the average distance $c_{int}$ between the conducting CuO$_{2}$
planes, one can obtain $n_{s2d}/m^{*}$ where $n_{s2d}$ denotes
the 2-dimensional carrier density.  We remind here that the Fermi energy
$\epsilon_{F}$ of 2-dimensional electron gas is proportional to
$n_{n2d}/m^{*}$.  Thus, we can consider $n_{s2d}/m^{*}$ as a parameter
representing kinetic energy for translational motion of superconducting
carries.  For 3-d systems without phase separation,
in which $n_{n}=n_{s}$, one 
can combine the muon relaxation rate $\sigma \propto n_{s}/m^{*}$ with 
the Sommerfeld
constant $\gamma(T=T_{c})\propto n_{n}^{1/3}m^{*}$, or 
Pauli susceptibility $\chi(T=T_{c}) \propto n_{n}^{1/3}m^{*}$, 
to deduce $\epsilon_{F} \propto n_{s}^{2/3}/m^{*}$.
The ``effective'' Fermi energy $\epsilon_{F}$ obtained in this way 
does not necessarily correspond to the real Fermi energy 
in band structure calculations.  Since $1/\lambda^{2}$ corresponds to  
the Drude spectral weight in optical conductivity which
condenses into a delta function at $\omega = 0$ below $T_{c}$, 
$\epsilon_{F}$ might also be called as a Drude energy scale.

Figure 3 shows a plot of $T_{c}$ versus $T_{F} = \epsilon_{F}/k_{B}$
thus obtained from the results of $n_{s}/m^{*}$ [7].
For A$_{3}$C$_{60}$ where a reliable value of {\it l\/} is
not available, we used the clean-limit value of $n_{s}/m^{*}$
without corrections regarding $\xi$/{\it l\/} [17,18].  
We find that HTSC, organic BEDT [19], and some other systems [17-20] have
very high and nearly equal ratios of $T_{c}/T_{F}$.
The $T_{B}$ line in this figure shows the Bose-Einstein 
condensation temperature for a non-interacting 3-d Bose gas
having the boson density $n_{B}= n_{s}/2$ and mass $m_{B}=2m^{*}$.
Compared to $T_{B}$, the actual transition temperatures 
$T_{c}$ of HTSC systems are reduced by a factor of 4 to 5.
This reduction is natural in view of overlapping
pairs which would reduce $T_{c}$, and in view of 2-dimensional
character of HTSC systems.  However the parallel behavior of
$T_{B}$ and the observed results of $T_{c}$ suggests that
the origin of the linear relationship between $T_{c}$  
and $n_{s}/m^{*}$ could be deeply related to BE condensation.
Figure 3 serves as an empirical way to classify various 
superconductors in a crossover from BE to BCS condensation.

\section{HTSC Systems involving microscopic heterogeneity: Zn-doping and 
overdoping}
 
\subsection{(Cu,Zn) substitution}

When a very small amount of Zn is substituted for Cu on the
CuO$_{2}$ planes of HTSC systems, $T_{c}$ is reduced.
Figure 4(a) shows the reduction of $\sigma(T\rightarrow 0)$
as a function of Zn concentration $c$ substituting in-plane Cu.
The superfluid density $n_{s}/m^{*}$ decreases with increasing
$c$.  To account for this reduction, we proposed ``Swiss Cheese
Model'' [9] where each Zn is assumed to destroy superconductivity
in the surrounding region with the area $\pi\xi_{ab}^{2}$, as 
illustrated in Fig. 4(a). 
The solid lines in Fig. 4(a), which represent predictions of this model,
are obtained just from $c$ and $\xi_{ab}$ deduced from the upper critical
field $H_{c2}$, {\sl without any fitting\/} to the data.  
Sufficiently good agreement
between these lines and the observed results indicate that
carriers within the $\pi\xi_{ab}^{2}$ area around each Zn no longer
contribute to the superfluid density $n_{s}/m^{*}$.  
Subsequently, this picture was directly confirmed by Pan et al. [21] in 
Scanning Tunneling Microscopy studies, where the local density of 
states around Zn showed features characteristic to normal 
regions, as shown in Fig. 4(b).  These $\mu$SR and STM 
results are consistent with the non-zero $\gamma$ term
of the specific heat, which increases with 
increasing Zn concentration [22].  In the plot of $\sigma$ versus $T_{c}$
in Fig. 1, the results of Zn-doped systems (closed triangle and star
symbols) follow the trajectory of 
simple hole doped HTSC systems (open symbols).

\subsection{overdoping} 

Tl$_{2}$Ba$_{2}$CuO$_{6+\delta}$ (Tl2201) systems have been 
extensively studied by various experimental methods as a
prototype of overdoped HTSC systems.  Tl2201 compounds have very 
small residual resistivity, which assures that the system lies well within
the clean limit.  To our surprise, with increasing overdoping
from the nearly optimal $T_{c}$ = 85 K sample, the relaxation rate
$\sigma(T\rightarrow 0)$ in TF-$\mu$SR decreased [8,15] as shown by the 
closed circle symbols in Fig. 1.  
This behavior is also shown in a plot of $\sigma(T\rightarrow 0)$ versus
doping factor $\delta$ in Fig. 5(a).  
In view of no anomaly in $m^{*}$,
the different behaviors of the normal state carriers $n_{n}$
and the superfluid density $n_{s}/m^{*}$ suggest
that carriers are spontaneously separated into those which are involved
in superconductivity and those which remain unpaired fermions, as illustrated 
in Fig. 5(a).  

By defining the ``gapped'' and ``ungapped'' responses
in the specific heat measurements, as illustrated in Fig. 5(a),
we plotted the $\delta$ dependence of the gapped response obtained
from the data of Loram {\it et al.\/} [22] in Tl2201 in Fig. 5(a).  
The good agreement between the
$\mu$SR superfluid density $\sigma(T\rightarrow 0)$ (closed circles) and the 
``gapped'' response in specific heat (open symbols) provides a
support to our view [8,23]
with spontaneous phase separation between superconducting and normal
regions in overdoped HTSC systems.  

The volume fraction of
the superconducting region can also be estimated from the 
``specific heat jump'' $\Delta C$.  In BCS superconductors
$\Delta C \propto C_{n} \propto \gamma_{n} T$, where $C_{n}$ denotes the
normal-state specific heat, and $\gamma_{n}$ stands for the Sommerfeld     
constant derived in the normal state.  
Then, in the plot of $C/T$ versus $T$ as in the inset of Fig. 5(a),
one would expect that the specific heat jump $\Delta C/T$
should be independent of $T_{c}$, for systems having a common
$\gamma_{n}$ value.  In Loram's specific heat results of Tl2201 [22],
$\gamma_{n}$ does not depend on doping, while the jump
$\Delta C/T$ decreases with increasing doping.  This 
observation further supports decreasing superfluid volume fraction
with increasing doping (decreasing $T_{c}$) in Tl2201.

Residual normal 
response in the overdoped cuprate was also found in 
measurements of optical conductivity [24].
Furthermore, we calculated an expected doping dependence 
of $n_{s}/m^{*}$ based on 
a simple model assuming phase separation [23] (as described in
section VII), and obtained a good agreement with the observed results
in (Y,Ca)Ba$_{2}$Cu$_{3}$O$_{y}$ [25], as shown in Fig. 5(b).
So far, there is no direct
observation reported regarding the size of these phase-separated regions
in overdoped HTSC.    
 
\section{HTSC systems with static SDW nano-islands}    

Magnetic order of the parent compound La$_{2}$CuO$_{4}$ of the 
214 cuprates was first confirmed by ZF-$\mu$SR measurements [26].
Figure 6(a) shows the time spectra obtained in the ZF measurements 
of antiferromagnetic
La$_{2}$CuO$_{4}$ (AF-LCO) which has the N/'eel temperature $T_{N} > 250$ K.
As in most other magnetic systems, we find that muon spin precession 
sets in below $T_{N}$, with the precession amplitude 
independent of temperature, while the frequency increasing with 
decreasing temperature as the sub-lattice
magnetization builds up.  In the La214 systems with the hole concentration 
near 1/8 per Cu, incommensurate static spin correlations have
been found by neutron scattering [27].  Time spectrum of ZF-$\mu$SR 
in La$_{1.875}$Ba$_{0.125}$CuO$_{4}$ (LBCO:0.125) [28] is shown in 
Fig. 6(b).  In this case, we find a Bessel function line shape
characteristic of ZF-$\mu$SR in incommensurate magnetic systems [29,30],
such as the one observed in (TMTSF)$_{2}$PF$_{6}$ [29].

La$_{2}$CuO$_{4.11}$ (LCO:4.11) is a system with oxygen intercalated
in a stage-4 structure.  This system is superconducting with $T_{c}$
= 42 K, which is the highest among the La214 family, while also 
exhibiting static incommensurate magnetism below $T_{N}$ = 42 K [31]
This incommensurate modulation has a very long correlation length
($\geq 600$ \AA), as determined from a very sharp satellite magnetic
Bragg peak in neutron scattering results.  ZF-$\mu$SR spectra of 
this system [10,32], shown in Fig. 6(c), has a 
Bessel-function line shape, as expected for an incommensurate spin structure.
The amplitude of this precession, however, increases gradually
below $T_{N}$ with decreasing temperature, while the frequency is
almost independent of temperature below $T_{N}$.  Furthermore,
the amplitude of precessing signal at $T\rightarrow 0$ 
is less than half of that in 
LBCO:0.125 (Fig. 6(b)), which indicates that the static magnetism 
exists in less than a half of the total volume.   

In Fig. 7(a) and (b), we show temperature
dependences of the volume fraction $V_{M}$ of muons 
in the region with static magnetic freezing, derived from the
precession amplitude, and the frequency $\nu$ observed in 
LCO:4.11, (La$_{1.88}$Sr$_{0.12}$)CuO$_{4}$ (LSCO:0.12),
LBCO:0.125 and (La$_{1.475}$Nd$_{0.4}$Sr$_{0.125}$)CuO$_{4}$
(LNSCO:0.125) [33]. 
In LCO:4.11 and LSCO:0.12, where superconductivity coexists
with static magnetism, the static 
magnetism is confined to a partial volume fraction.
As shown in Fig. 7(c), the temperature dependence of 
the neutron Bragg intensity in LCO:4.11 is consistent 
with the behavior of $V_{M}\nu^{2}$.  Unlike usual magnetic systems,
however, the $T$ dependence is mostly due
to the change of the volume fraction $V_{M}$.

The internal field at the muon site in HTSC systems is due 
to dipolar field from neighbouring static Cu spins.  
Depending on the range of this field, the volume fraction of 
muons $V_{M}$ subject to static field is somewhat larger than
the volume fraction $V_{Cu}$ of frozen Cu spins.  Figure 8(a)
shows our simulation results for the relationship between
$V_{Cu}$ and $V_{M}$ for cases with incommensurate static 
magnetism in island regions having a radius $R$ [10]. 
Here we see that about 70-80\%\ population
of static Cu spins is enough to create observable static fields
at the sites of all the muons. Thus, our previous result [33] on 
La$_{1.45}$Nd$_{0.4}$Sr$_{0.15}$CuO$_{4}$ (LNSCO:0.15), 
which is a superconductor with $T_{c} \sim 10 K$ with $V_{M} > 95$\%\,
is still compatible with a picture that superconductivity and static
magnetism occur in mutually exclusive regions of CuO$_{2}$ plane.

Although the relationship between $V_{M}$ and $V_{Cu}$ is 
nearly independent of the radius $R$ of static SDW islands,
the damping rate $\Lambda$ of the Bessel oscillation 
depends on $R$, as shown in Fig. 8(b).  By converting
observed $V_{M}$ into $V_{Cu}$ using Fig. 8(a), and then
plotting observed $\Lambda$ in Fig. 8(b), we find  
that our ZF-$\mu$SR results in LCO:4.11 are consistent
with the island size $R \sim$ 15 - 30 \AA [10].  Note
that this length scale is comparable to $\xi_{ab}$.
This model of ``static SDW nano islands'' can be reconciled
with the long range spin correlations found by
neutron Bragg peaks, if we consider percolation of these islands
via a help of interplaner correlations,
as proposed in [10].

To study the relationship between the magnetic volume fraction
$V_{Cu}$ and the superfluid density $n_{s}/m^{*}$, we performed
$\mu$SR measurements in (La$_{1.85-y}$Eu$_{y}$Sr$_{0.15}$)CuO$_{4}$
(LESCO) [11,34] where the concentration of doped hole carriers
is fixed to 0.15 per Cu, while the magnetic volume changes
with increasing Eu concentration $y$.  
The volume fraction with static magnetism was 
determined by the ZF-$\mu$SR, while $n_{s}/m^{*}$ was
measured in TF-$\mu$SR.
The results shown in Fig. 9 clearly demonstrate a
trade-off between $V_{Cu}$
and the superfluid density $n_{s}/m^{*}$.  This implies that
superconductivity does not exist in the volume which has static 
magnetism.
  
In the plot of $\sigma(T\rightarrow 0)$ versus $T_{c}$ in Fig. 1,
the results of these systems with static stripe magnetism, shown
by the striped square symbols, 
follow the trajectory of simple hole-doped and Zn-doped
systems.  This indicates that $n_{s}/m^{*}$ is again a determining
factor of $T_{c}$ in HTSC systems where superconductivity
and magnetism coexist.  We also note that the results in overdoped Tl2201
can be viewed as following a monotonic relationship between $T_{c}$
and the superfluid density, with the slope roughly comparable to 
cuprates with/without other types of perturbations.

\section{Analogy with superfluid He films}

Superfluid He films represent another set of systems where 
$T_{c}$ is strongly correlated with the 2-dimensional superfluid
density $n_{s2d}$ at $T\rightarrow 0$.  Using the published results, 
we made Fig. 10 [35] which shows  
correlations between $T_{c}$ and $n_{s2d}/m^{*}$
for $^{4}$He films on Mylar substrate (open circle symbol) [36], 
on Vycor Glass which represents a porous media (star symbol) [37,38], 
as well as for $^{4}$He/$^{3}$He mixtures adsorbed on fine
alumina powders (closed diamond symbol) [39].  The results on Mylar films show
linear relationship, consistent with the behavior expected for
the KT transition [40] (see next section for details) as indicated
by the solid line.  With the lowest level of perturbation, 
this case corresponds to simple hole-doped HTSC systems
in comparison between the cuprates and He films.
The results on Vycor Glass is analogous to the Zn-doped
cuprates: in both cases some normal regions are formed
as a ``healing region/layer'', while $T_{c}$ is still strongly 
correlated with the superfluid density despite non-trivial
geometry of the superfluid.     

Mixture of bosonic $^{4}$He and fermionic $^{3}$He liquids exhibits
a phase diagram shown in the inset of Fig. 10.  With increasing
$^{3}$He fraction, $T_{c}$ decreases, being roughly proportional
to the volume fraction of $^{4}$He.  In bulk geometry, the mixture
undergoes macroscopic phase separation into $^{4}$He-rich superfluid and 
$^{3}$He-rich normal fluid, the heavier superfluid 
existing underneath the lighter normal fluid in a container.
This phase separation can be confined into a microscopic length
scale by adsorbing the mixture onto porous media or fine powders [39,41],
where superfluidity remains up to a large $^{3}$He
fraction.  The results for the $^{4}$He/$^{3}$He mixture
in Fig. 10 exhibit a behavior very similar to that of 
overdoped Tl2201 in the cuprates in Fig. 1.  In both the He mixture
and the overdoped HTSC, inclusion of too many 
fermions (doped holes and $^{3}$He)
results in microscopic phase separation and reduction of 
both $T_{c}$ and the superfluid density.  Thus, we find a
remarkable similarities [35] in the plot of $T_{c}$ versus
superfluid density in cuprates (Fig. 1) and He films (Fig. 10)
for the cases with/without perturbation.

Figure 1 shows correlations between $T_{c}$ and 3-dimensional (3-d)
superfluid density while Fig. 10 shows $T_{c}$ versus
2-d superfluid density.  By knowing the average interlayer
distance $c_{int}$ between the CuO$_{2}$ planes, we can 
generate a plot of $T_{c}$ versus $n_{s2d}/m^{*}$ for 
the cuprate systems.  The resulting plot, Fig. 11, clearly indicates
that in HTSC systems, $T_{c}$ is higher for shorter $c_{int}$
for a given 2-d superfluid density $n_{s2d}/m^{*}$ [42].  
This observation is consistent with the variation of $T_{c}$
reported for multilayer films with alternating layers of 
superconducting YBa$_{2}$Cu$_{3}$O$_{7}$ (YBCO) and insulating
PrBa$_{2}$Cu$_{3}$O$_{7}$ (PBCO) [43], as shown in the inset of Fig. 11.
These results indicate that the interlayer
coupling is essential for obtaining higher $T_{c}$ in HTSC systems.
The observed variation of $T_{c}$  
cannot be explained by the simplest version of KT transition where
$T_{c}$ should be determined solely by $n_{s2d}/m^{*}$.
   
\section{Condensation mechanisms}

\subsection{BE and BCS condensation and KT transition}

Correlations between $T_{c}$ and the superfluid density
shown in Fig. 1 and Fig. 11 help consideration of condensation
mechanisms in HTSC systems.  In BE condensation [44], the bosons are 
pre-formed at a very high ``paring'' temperature $T_{p}$.
At a given temperature $T < T_{p}$ each boson has kinetic
energy of $k_{B}T$, which defines the thermal wave length
$\lambda_{th}$ representing the 
spread of the wave function of this boson due to the uncertainty
principle.  With decreasing $T$, $\lambda_{th}$ increases.
When $\lambda_{th}$ becomes comparable to the inter-particle distance
$n_{B}^{-1/3}$, the wave functions of neighbouring bosons 
start to overlap, as illustrated in Fig. 12(a).  
Thanks to the tendency of bosons to fall into the
same state by building up phase coherence of wave functions,
Bose condensation occurs at $T$ where $\lambda_{th} \sim n_{B}^{-1/3}$.
From this, we have the BE condensation temperature $T_{B} \propto
n_{B}^{2/3}/m_{B}$, where $n_{B}$ and $m_{B}$ denote the density
and mass of the boson.  In this way, the superfluid density and $T_{c}$
is directly related in BE condensation.

In BCS condensation [45], the energy scale of attractive 
interaction determines the energy gap and $T_{c}$.  
The effective Fermi energy 
$\epsilon_{F} \propto n_{n}^{2/3}/m^{*}$ is much larger than
$kT_{c}$ in BCS condensation, i.e., there is a sufficient
density of fermions above $T_{c}$.  However, the 
system stays in the normal state until the temperature is reduced to 
the pairing energy scale $T_{p}$, where bosons are formed.  
Once pairs are formed,
their boson density is high enough at $T_{c}$, and thus
the condensation occurs immediately at $T_{c} \sim T_{p}$.
In BCS condensation, $T_{c}$ depends on carrier density through the 
density of states at the Fermi level which determines 
the strength of electron-phonon interaction.  
However, this dependence is rather indirect.
Suppose we have a BCS superconductor with the 
carrier density $n_{s}$.  If the Debye frequency 
is doubled, $T_{c}$ and 
the energy gap $\Delta$ would be doubled.  However
the carrier density stays unchanged, as can be found from 
the illustration of Fig. 12(b). This example
shows that the superfluid density is not a direct
determining factor for $T_{c}$ in BCS condensation.

Kosterlitz-Thouless transition [40] is a phenomenon in 
2-dimensional superfluids / superconductors.  
The elementary excitation of superfluid He film is the formation of
vortex anti-vortex pairs.  The energy required for this
is governed by the phase stiffness, which is proportional
to the superfluid density.  At the transition temperature
$T_{KT}$, the thermal energy becomes sufficient to form
un-bound vortices, which results in dissipation and destruction of 
superfluidity / phase coherence.  Thermodynamic arguments lead to the 
universal relationship between the 2-d superfluid density at
$T=T_{KT}$ and the value of $T_{c}$ as
$kT_{KT}\propto n_{s2d}/m^{*}(T=T_{KT})$, which should be
independent of material.  This relationship was confirmed
as the jump of the superfluid density at the superfluid transition of
He films [46].

\subsection{Evolution from 3-d to 2-d}

Figure 13(a) illustrates evolution of various energy
scales in superfluid He film and a thin film of BCS superconductor. 
The BE condensation temperature of He in the 3-d limit 
corresponds to the lambda temperature $T_{\lambda} = 2.2$ K, 
which is close to $T_{B}\propto n_{B}^{2/3}/m^{*}$, except for
some reduction due to departure of real system from an 
ideal non-interacting Bose-gas limit.  
With decreasing film thickness $d$, $T_{c}$ starts to change when the thickness
becomes comparable to a few times the coherence length 
$\xi \sim n_{B}^{-1/3}$.  For $d \leq \xi$,
$T_{c} \propto n_{s2d}/m^{*}$.  In superfluid He film on Mylar, 
the superfluid density at $T\rightarrow 0$ is very close to that at just below
$T_{c}$, as shown in Fig. 13(a).  
Thus the Kosterlitz-Thouless relation holds for 
$n_{s2d}$ at $T\rightarrow 0$, as can be seen in Fig. 10.
Between $T_{c}$ and $T_{\lambda} = 2.2$ K, 
the order parameter exhibits phase fluctuations, i.e., 
superfluidity is dynamic.    
The pair formation energy scale is much higher than 
$T_{c}$ in He for any thickness value $d$.

In a thin film of a BCS superconductor, $T_{c}$ should show a similar
reduction from the value $T_{c3d}$ in the bulk 3-d system, 
when the film thickness becomes comparable to
or smaller than a few times $\xi$ in the clean limit (and 
{\it l\/} in the dirty limit), as shown in 
Fig. 13(b).  The superfluid density 
$n_{s}/m^{*}$ at $T\rightarrow 0$ 
would follow a thickness dependence similar to that of $T_{c}$.
Note that $n_{s}/m^{*}$ is very large, 
as it is related to $T_{F} \propto n_{n}^{2/3}/m^{*} \gg T_{c}$
in the 3-d limit.  There should be difference between $n_{n}/m^{*}$
and $n_{s}/m^{*}$ in the region $d < \xi$ (or {\it l\/}), due to difficulty 
in forming pairs in a restricted geometry.
On the other hand, the general argument of KT
relationship should still hold.  This is possible only
if $n_{s2d}/m^{*}$ at $T=T_{c}$ for $d < \xi$ is much smaller
than $n_{s2d}/m^{*}$ at $T\rightarrow 0$, as illustrated
in Fig. 13(b).   Therefore, the ``jump'' of the superfluid density
at $T_{KT}$ is practically invisible in a BCS thin film.
For $d \leq \xi$, the region $T_{c} < T < T_{c3d}$ is
characterized by phase fluctuations.  However, the 
pair formation occurs at $T \sim T_{c3d}$ for any thickness.
Consequently, bosons exist only below $T_{c3d}$, as illustrated
in Fig. 13(b). 

\subsection{BE-BCS Crossover and Phase Fluctuation Models}
  
In 1989 [6] and 1991 [7], we suggested the
relevance of BE condensation to the universal relationship between
$T_{c}$ and $n_{s}/m^{*}$ in HTSC systems shown in Fig. 1 [6].
By combining this phenomenon with the pseudogap behavior 
then observed by NMR and conductivity, 
we proposed a picture of BE to BCS crossover in 1993-94 [47,48], as  
illustrated in Fig. 14(a) [42,47,48].
General concept of the BE-BCS crossover has been considered
earlier by several scientists [49], including Randeria
and co-workers [50] who adopted this concept to the interpretation 
of the susceptibility/NMR results for the pseudo gap.
Within our knowledge, however, our proposal was the first to 
combine the results of superfluid density with the pseudogap behavior
in the underdoped cuprates.  As shown in Fig. 14(a), we regarded  
the pseudogap temperature $T^{*}$
as the pair formation temperature $T_{p}$.  
The reduction of the c-axis dc conductivity below $T^{*}$ can be 
interpreted as resulting from reduced interplaner tunneling
probability for paired bosonic carriers having 2e charges,
while reduction of magnetic susceptibility can be attributed to
the formation of a spin singlet bosonic (pairing) state in this picture.

In 1995, Emery and Kivelson (EK) [51] proposed a model based on phase
fluctuations, as illustrated in Fig. 14(b).  The 
BE-BCS model and the phase fluctuation model
share many features in common.  However, there are several
important differences.  Based on the linear
relationship between $T_{c}$ and the superfluid
density at $T\rightarrow 0$, EK presented arguments
basically parallel to that for the KT transition in 2-d systems.
They pointed out that $T_{c}$ of the underdoped cuprates
is determined by the energy scale for the phase fluctuations.
In their picture, the entire region of pseudogap below $T^{*}$ 
is characterized by phase fluctuations.  They argued that
the 2-dimensional aspect is essential for obtaining high $T_{c}$.      

In order to calculate $T_{\theta}^{max}$ which denotes the energy scale 
for phase fluctuations to destroy superconductivity, 
EK multiplied $\lambda^{-2} \propto n_{s}/m^{*}$
to the interplaner distance $c_{int}$ for HTSC and some other 
2-d systems.  This is equivalent of obtaining $n_{s2d}/m^{*} \propto
T_{F}$ in 2-d. 
For 3-d systems, EK multiplied $n_{s}/m^{*}$ and the
coherence length $\xi$,
which leads to an energy scale much higher than $T_{F}$.
This energy scale is unrealistically high,
and irrelevant to condensation arguments.
If one substitutes the interparticle distance 
$n_{s}^{-1/3}$ instead of $\xi$, we recover
$T_{F}$ for 3-d systems.  Then, Table 1 in ref [51] by EK becomes
essentially equivalent to Fig. 3 of our 1991 paper [7],
shown as a part of Fig. 3 of this article, where we plotted $T_{c}$
versus $T_{F}$.   
We consider that the distinction between 2-d and 3-d systems by EK 
is an artefact resulting from the overestimate of $T_{\theta}^{max}$
in 3-d systems.
In general, the 2-dimensional aspect does not help increasing $T_{c}$
of any superconductor, as can be found in Fig. 13. 

Figure 15 shows the current estimates of $T^{*}$ from various
methods as plotted versus hole concentration.  
If the entire region below $T^{*}$ is supposed to have 
superconducting phase fluctuations, as conjectured by 
EK, the situation is rather
similar to a thin-film BCS superconductor shown in Fig. 13(b).
In contrast, if $T^{*}$ is solely representing the pair formation,
as proposed in our BE-BCS crossover picture, and if there is 
a 2-dimensional aspect remaining in HTSC systems, there should be
two different energy scales above $T_{c}$,
as we discussed in ref [35].  They are the 
temperature $T_{dyn}$ at which the phase coherence of bosons 
completely disappear 
and dynamic superconductivity vanishes, in addition to 
the temperature $T_{p}$ at which pairs are dissolved into fermions. 
This situation is similar to the case of He films shown in Fig. 13(a).
One of important differences between BE-BCS and EK conjectures
lies in this point.  
 
\subsection{Pair formation and dynamic superconductivity}

Since high-$T_{c}$ cuprates have a highly 2-dimensional electronic
structure, several experiments have detected the ``dynamic superconductivity''
existing above $T_{c}$:
(1) In YBCO-PBCO films with a thick PBCO layers separating a single layer
YBCO, $T_{c}$ is reduced to 15-20 K.  Between this temperature and
the $T_{c} \sim 90$ K of the bulk YBCO, one observes reduction of 
the normal state conductivity, which follows the predictions of the 
Kosterlitz-Thouless theory [52].
(2) Corson {\it et al.\/} [53] measured dynamic superfluid response
in underdoped Bi2212, and found that the response
depends on the measuring frequency $\omega$ above a certain
``branch-off'' temperature $T_{off}$, as shown in Fig. 16.  Furthermore, 
the superfluid density 
$n_{s}/m^{*}$ at $T=T_{off}$ agrees well with the 
universal value $n_{s}/m^{*}(T=T_{KT})$
expected for the KT transition.
The critical temperature $T_{c}$, at which $n_{s}/m^{*} =0$, 
increases with increasing $\omega$,
suggesting that this phenomenon corresponds to 
the ``dynamic superfluidity'' expected above $T_{c}(\omega=0)$
and below $T_{dyn}$.  It should be noted, however, that 
$T_{dyn}$ for the measuring frequency of 600 GHz is limited
to $T\leq 100$ K.

These results indicate that certain cuprate systems with
high anisotropy,
such as underdoped Bi2212 and single-layer YBCO film, exhibit
dynamic superconductivity as
expected from the KT theory.  However, these results
do not necessarily provide explanation to the origin of the pseudo gap
at $T=T^{*}$, since $T_{dyn}$ observed in Bi2212 is significantly
lower than the pseudo-gap temperature $T^{*}$.
Namely, the phase fluctuations alone
cannot explain the entire pseudo-gap phenomena.

Dependence to an external magnetic field should be quite
different if these two distinct energy scales above $T_{c}$ 
correspond respectively to pairing and dynamic superconductivity.
The former would not depend much on external fields, while the
latter should be very sensitive to the field.  
Indeed, the pseudo gap below $T^{*}$ was found to be 
insensitive to the applied field in tunneling measurements [54].
Recently, Lavrov {\it et al.\/} [55] found that c-axis
conductivity shows negative magneto-resistance upon cooling at
a temperature well below $T^{*}$ but above $T_{c}$.  
This negative magneto-resistance may be due to 
the onset of dynamic superconductivity below $T_{dyn}$.

The results of the Nernst effect [56], which appear in the underdoped region of
La214 systems above $T_{c}$ but well below $T^{*}$ can also
be interpreted as possible evidence for dynamic superconductivity
below $T_{dyn}$.  Other phenomena potentially related to this
energy scale include: 
(a) superfluid response of the ARPES coherence
peak [57]; and (b) 41 meV neutron resonance mode [58].

Based on these considerations, we propose a new phase diagram,
with two distinct lines of $T^{*}$ and $T_{dyn}$
in the underdoped region as shown in Fig. 17.  Note that the pair formation
at $T^{*}$ is necessary for superconductivity in cuprates, but
is not enough to support any phase coherence.  It is only
when the thermal energy scale becomes less than the energy scale
representing the number density of bosons that the phase coherence
can set in.  The onset temperature $T_{dyn}$ of 
dynamic superfluidity represents this energy scale.  
In highly 2-dimensional cases, one should further
cool down below $T_{dyn}$ before achieving long-range phase coherence
at $T_{c}$.

\section{Evolution from superconductor to normal metal}

\subsection{phase diagram for the overdoped cuprates}

In BCS superconductors in the clean limit, all the carriers in the 
Fermi sphere contribute to superfluid.  There should be no 
normal carriers remaining at $T \rightarrow 0$.  Therefore,
neither the BE-BCS nor the phase fluctuation
pictures can explain the reduction of $n_{s}$ with 
increasing hole doping observed in several overdoped cuprate systems
[8,15,25] described in section III-B.  
 
Tallon, Loram and co-workers [59] have noticed that the $T^{*}$ line
may be heading towards $T=0$ at a ``critical hole concentration''
around $x_{c} \sim 0.19$.   $\mu$SR studies of Panagopoulos {\it et al.\/} 
[60] show that static magnetism, either spontaneously existing or
induced by Zn doping, disappears at $x \geq x_{c}$.  
Generally, increasing hole doping would tend to destroy magnetic 
interactions in the cuprates.

Let us make the following three assumptions:
(a) the paring in HTSC systems is due to a magnetic interaction;
(b) the $T^{*}$ line represents the pair formation, and
(c) the $T^{*}$ line disappears at $x = x_{c}$.  
Then, no genuine superconducting pairing exists
in the overdoped region at $x \geq x_{c}$.  However, if the energy loss
for charge disproportionation is overcome by the gain of  
pairing and condensation energies, the system can spontaneously
phase separate into a ``hole-poor'' superfluid with the local
hole concentration $x \leq x_{c}$ and a ``hole-rich'' normal fermion
region with $x > x_{c}$.  If this phase separation remains
microscopic with the length scale comparable to $\xi_{ab}$, 
we can expect superconductivity in the overdoped region,
similarly to the case of $^{4}$He/$^{3}$He mixture.

The energy loss of charge disproportionation can be estimated 
following Coulomb blockade type calculation and/or formation of nano-islands
serving as capacitor allays.  Our crude estimate [23] shows that
this energy becomes comparable to paring/condensation energy 
gain in the cuprates.  Based on these considerations, we proposed
a phase diagram shown in Fig. 17, with phase separation 
in the overdoped region.  As shown in Fig. 5(b), 
the doping evolution of 
the superfluid density $n_{s}$, calculated using this model [23],
shows a good qualitative agreement with the observed results.
We also note that this model can reproduce a very sharp temperature
dependence of 
$n_{s}/m^{*}$ and $H_{c2}$ at $T\rightarrow 0$ found in the heavily overdoped
region of Tl2201 [8,15,61].  

Tallon and Loram [59] presented a view that the
existence of superconductivity in the overdoped cuprates
at $x \geq x_{c}$, where the $T^{*}$ line
does not exist, implies that the magnetic
interaction below the $T^{*}$ line is not required for superconductivity  
but rather is a competing factor which weakens superconductivity.
The present model with phase separation provides an alternative
view where the pseudogap phenomena,
as a necessary factor (pair formation) for superconductivity,
can be compatible with superconductivity in the overdoped cuprates.

\subsection{analogous cases in other systems}

In HTSC cuprates, as well as in 2-d organic (BEDT-TTF)$_{2}$-X 
and 3-d A$_{3}$C$_{60}$
superconductors, superconductivity appears in the vicinity of 
the insulator to metal transition.  Superconductivity is taken over
by a presumably Fermi-liquid metallic state by
increasing carrier doping in cuprates, or by application of 
chemical and/or external pressure in BEDT and A$_{3}$C$_{60}$.
In all these cases, we expect that the normal state
spectral density (or Drude weight) $n_{n}/m^{*}$ to increase
in the process towards a simple Fermi liquid.  We showed 
that the superconducting spectral weight {\sl decreases\/} in 
this process for the case of cuprates.

We performed magnetization measurements of the in-plane
penetration depth in (BEDT-TTF)$_{2}$Cu(NCS)$_{2}$ under 
applied pressure [62], and found that $n_{s}/m^{*}$ decreases
with increasing pressure as shown in Fig. 18(a), contrary to 
the behavior of $n_{n}/m^{*}$ found in quantum oscillation
measurements [63].  This system is
known to be well in the clean limit.  These results indicate
that not only in cuprates but also in BEDT, the crossover from 
superconducting to metallic ground state is associated with
anomalous reduction of the superfluid spectral weight.

Similar behavior is also seen in A$_{3}$C$_{60}$ systems.
As shown in Fig. 1 and Fig. 18(b), the muon relaxation rate 
$\sigma(T\rightarrow0)$ measured in A$_{3}$C$_{60}$ decreases with
decreasing lattice constant [17,18], 
in the approach towards a simple metallic
state.  In view of rather large residual resistivity in the 
normal state, the results for these fullerides might be subject 
to correction related to the mean free path.  However, the
un-corrected raw data indicate that $1/\lambda^{2}$ again 
shows anomalous reduction when the system approaches 
presumably a simple metallic ground state.

These three systems exhibit similar
phase diagrams: superconductivity appears in the evolution from 
magnetic and insulating ground state to presumably simple
Fermi liquid state; in BEDT systems, a pseudo-gap
like behavior near the
magnetic phase was found in susceptibility.  
These features suggest a possibility
that the anomalous results in the overdoped cuprates may be a generic behavior
shared by a wider range of superconductors based on 
correlated electron systems.  We note that all these systems
follow the universal linear relationship in the plot of 
$T_{c}$ versus $n_{s}/m^{*}$, with approximately the same 
slope as shown in Fig. 1 and in ref. [62].
  
\section{Summary}

In this paper, we showed that $T_{c}$ in HTSC cuprate
systems exhibits universal correlations with 
the superfluid spectral weight $n_{s}/m^{*}$, for the
cases of simple hole doping, as well as for more
complicated cases with Zn-doping, overdoping and static
SDW nano island formation, where the system undergoes
spontaneous phase separation between superconducting
and normal regions with the length scale comparable to
the in-plane coherence length.  Robustness of this
relation for the case with/without perturbation
is analogous to the case of superfluid $^{4}$He
and $^{4}$He/$^{3}$He mixture films in non-porous and
porous media.  In all these cases, the superfluid 
density is the determining factor for $T_{c}$.
This is the basic feature of BE condensation.

By slightly revising the BE-BCS crossover picture, we have
proposed a new phase diagram for the cuprates which has
(a) two separate lines of $T^{*}$ and $T_{dyn}$ above
$T_{c}$ in the pseudogap state in the underdoped region,
(b) disappearance of the $T^{*}$ line at the critical hole
concentration $x_{c}$, and (c) phase separation in the overdoped
region.  In this model, the $T^{*}$ line represents pair
formation, whereas the $T_{dyn}$ line corresponds to the
onset of dynamic superconductivity.  Low dimensionality
prevents formation of long-range phase coherence 
between $T_{dyn}$ and $T_{c}$.   We also noted
anomalous reduction of the superfluid spectral weight
$n_{s}/m^{*}$ in the overdoped cuprates as well as in 
the 2-d organic BEDT and 3-d fulleride superconductors,
when the system approach simple metallic ground state.

\section{acknowledgement}

The author would line to acknowledge collaboration with
G.M. Luke, K.M. Kojima, S. Uchida, R.J. Birgeneau,
K. Yamada and discussions with O. Tchernyshyov and M. Randeria.
This work is supported by a grant from US NSF
DMR-01-02752 and CHE-01-17752.

\newpage

\begin{figure}

\begin{center}

\mbox{\psfig{figure=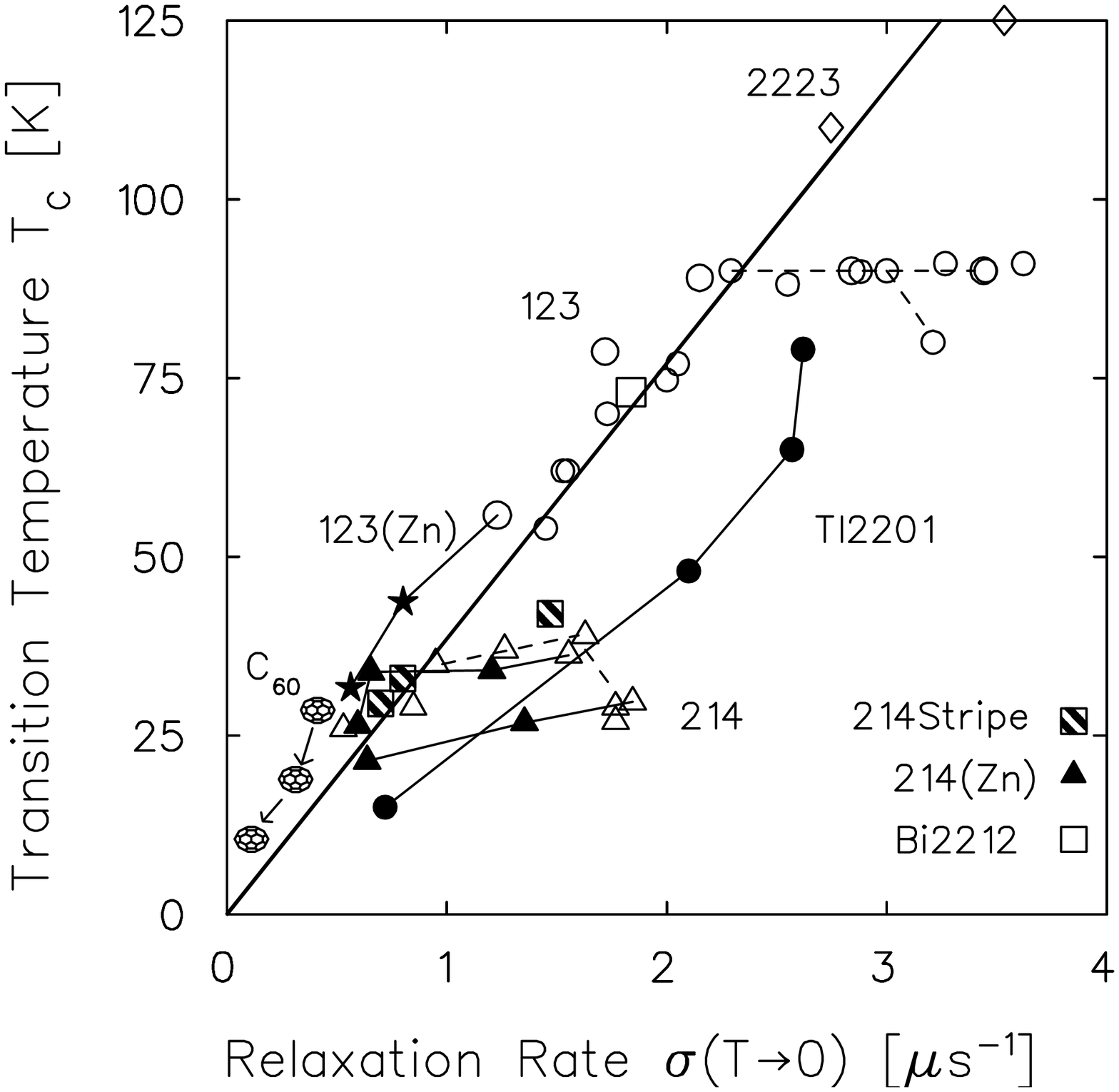,width=3in}}

\vskip 0.2 truein

\label{Figure 1.} 

\caption{A plot of the superconducting transition temperature $T_{c}$
versus the 
relaxation rate $\sigma(T\rightarrow 0)$ at low temperatures 
(proportional to the superfluid density
$n_{s}/m^{*}$) for several high-$T_{c}$ superconductors
[6-11], and fullerides A$_{3}$C$_{60}$ [17,18]. The results with the
``stripe square'' symbols represent points from
LESCO ([11]), 
LSCO:0.12, and LCO:4.11 ([10]) 
in the order of increasing $\sigma(T\rightarrow 0)$.
To account for difference between results for ceramic and single-crystal
specimens, the values for $\sigma$ for LCO:4.11 and LSCO:0.12,
observed with their c-axis parallel to the external field, 
were multiplied by 1/1.4.
}

\end{center}
\end{figure}
\newpage

\begin{figure}

\begin{center}

\mbox{\psfig{figure=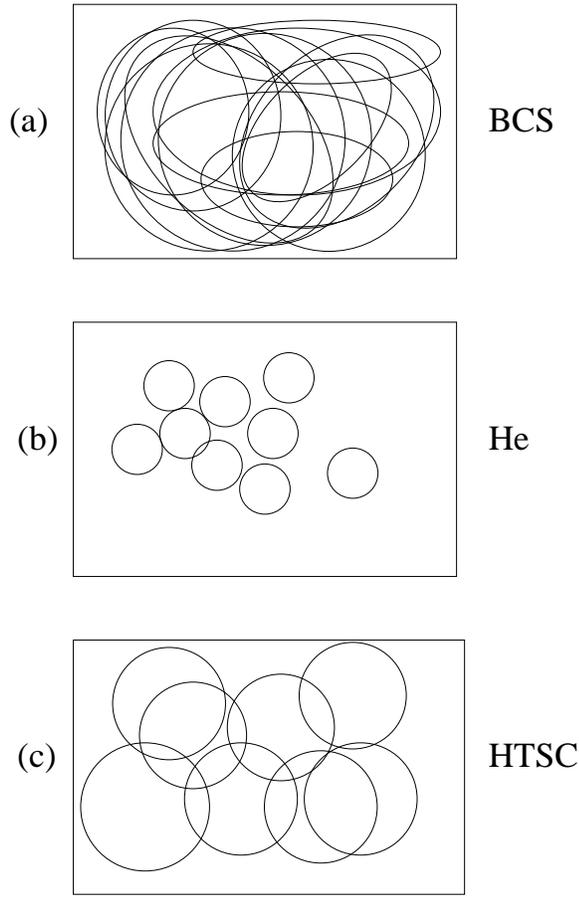,width=3in}}

\vskip 0.2 truein

\label{Figure 2.} 

\caption{Schematic view of overlap 
of pairs / bosons in (a) BCS superconductors,
(b) superfluid $^{4}$He, and (c) HTSC cuprate systems. The
circles represent the size of the pair, whose length scale is
given by the coherence length.  We estimate that about 3 to 5
pairs are overlapping with one another in the HTSC systems.
These results suggest that HTSC systems lie in crossover from 
BE to BCS condensation.}
 
\end{center}

\end{figure}

\newpage

\begin{figure}

\begin{center}

\mbox{\psfig{figure=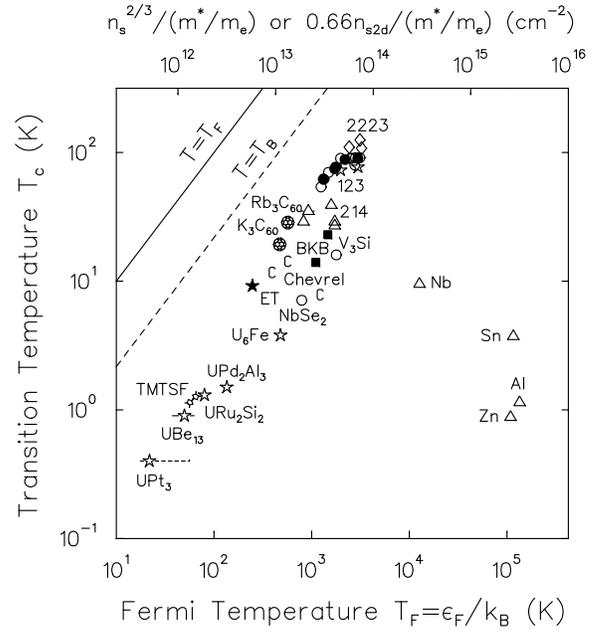,angle=90,width=3in}}

\vskip 0.2 truein

\label{Figure 3.} 

\caption{A plot of $T_{c}$ versus the effective Fermi temperature
$T_{F}$ obtained from the muon spin relaxation rate
$\sigma(T\rightarrow 0)$ (in combination with 
the average interlayer distance $c_{int}$ for 2-d systems, and
with the Sommerfeld constant $\gamma$ for 3-d systems) 
in various superconductors [7]. The broken line 
denotes Bose-Einstein condensation temperature $T_{B}$ 
for a given boson density $n_{s}/2$ and mass $2m^{*}$.}

\end{center}

\end{figure}

\newpage

\begin{figure}

\begin{center}

\mbox{\psfig{figure=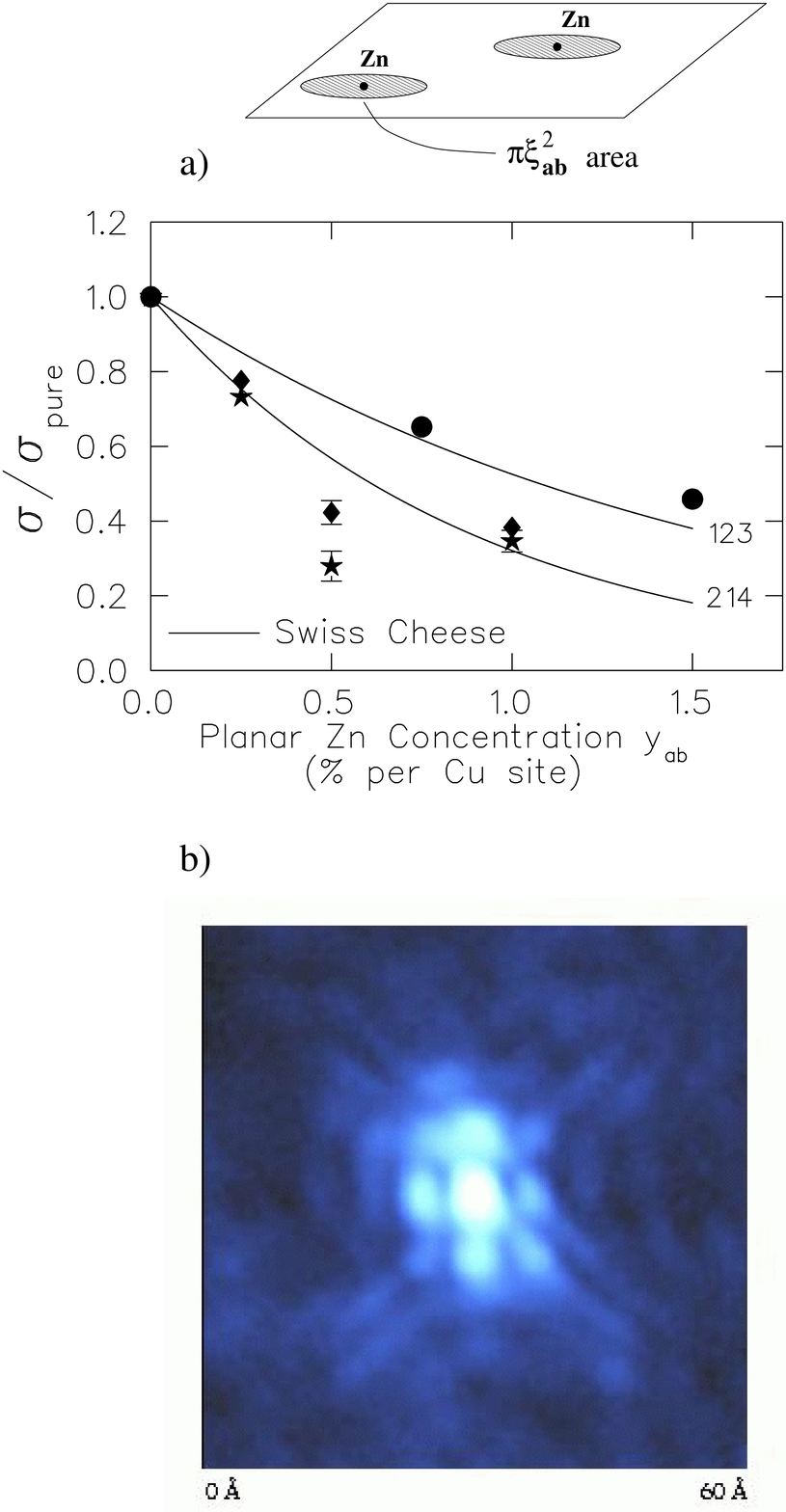,width=3in}}

\vskip 0.2 truein

\label{Figure 4.} 

\caption{(a) Muon spin 
relaxation rate $\sigma \propto n_{s}/m^{*}$
at $T\rightarrow 0$
plotted as a function of 
Zn concentration on the CuO$_{2}$ plane in 
YBa$_{2}$(Cu,Zn)$_{3}$O$_{6.63}$ (123) (closed circle), 
La$_{1.85}$Sr$_{0.15}$(Cu,Zn)O$_{4}$ (closed diamond),
and La$_{1.8}$Sr$_{0.2}$(Cu,Zn)O$_{4}$ (star symbol),
together with the estimate of ``Swiss Cheese model''
illustrated in the top figure [9].  In this model,  
carriers in the shaded region around each Zn impurity
do not contribute to the superfluid.
(b) The local density of states (LDOS) profile obtained by
Scanning Tunneling Microscope by Pan {\it et al.\/}
in Bi2212 around a Zn impurity [21].  The bright spot
in the center corresponds to high density of states
in the normal region, while the dark profile 
are from superconducting region with a gap in LDOS.}  

\end{center}

\end{figure}

\newpage

\begin{figure}

\begin{center}

\mbox{\psfig{figure=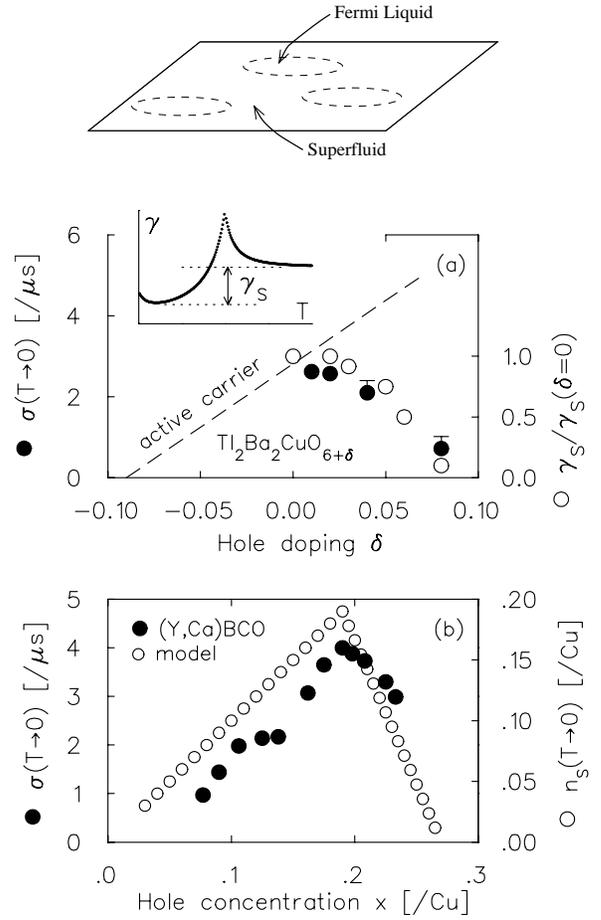,width=3in}}

\vskip 0.2 truein

\label{Figure 5.} 

\caption{(a) Muon spin relaxation rate $\sigma(T\rightarrow 0)\propto 
n_{s}/m^{*}$
(closed circles) [8] and the ``gapped'' response $\gamma_{s}$ 
in the linear-term of the specific heat (open circles) [22]
as a function of oxygen
concentration in Tl$_{2}$Ba$_{2}$CuO$_{6+\delta}$ (Tl2201).  The broken
line illustrates a projected variation of $n_{n}/m^{*}$.
(b) Doping dependence of $\sigma(T\rightarrow 0)\propto 
n_{s}/m^{*}$ observed in (Y,Ca)Ba$_{2}$Cu$_{3}$O$_{y}$ by
Bernhard {\it et al.\/} [25] (closed circles) and 
the superfluid density $n_{s}(T\rightarrow 0)$
calculated for a simple model with microscopic phase separation 
in overdoped cuprates [23], as illustrated in the top figure.}

\end{center}

\end{figure}

\newpage

\begin{figure}

\begin{center}

\mbox{\psfig{figure=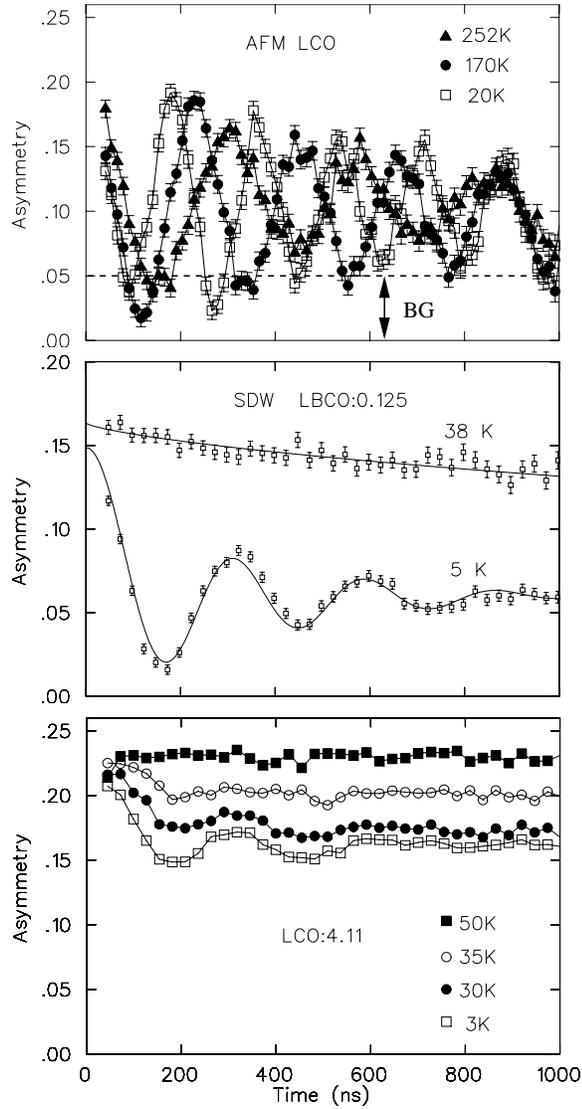,width=3in}}

\vskip 0.2 truein

\label{Figure 6.} 

\caption{Time spectra of ZF-$\mu$SR measurements in (a) 
antiferromagnetic La$_{2}$CuO$_{4}$ (AF-LCO),
(b) La$_{1.875}$Ba$_{0.125}$CuO$_{4}$ (LBCO:0.125)
and (c) stage-4 La$_{2}$CuO$_{4.11}$ [10].
In AF-LCO, the oscillation amplitude below $T_{N} > 250$ K
does not depend on $T$, while the frequency increases with
decreasing $T$, reflecting build-up of sub-lattice magnetization.
This is the behavior observed by $\mu$SR in most of uniform
ferro or antiferromagnetic systems.
In LBCO:0.125 [28], we see a Bessel function form at $T=5$ K, 
which is expected for the system with 
incommensurate static magnetism developed in the full volume
fraction.  In LCO:4.11, the amplitude of the oscillating 
and relaxing signal varies with $T$ without much change
in the frequency. The volume fraction of regions with static
magnetism is less that a half even at $T\rightarrow 0$ [10].}

\end{center}

\end{figure}

\newpage

\begin{figure}

\begin{center}

\mbox{\psfig{figure=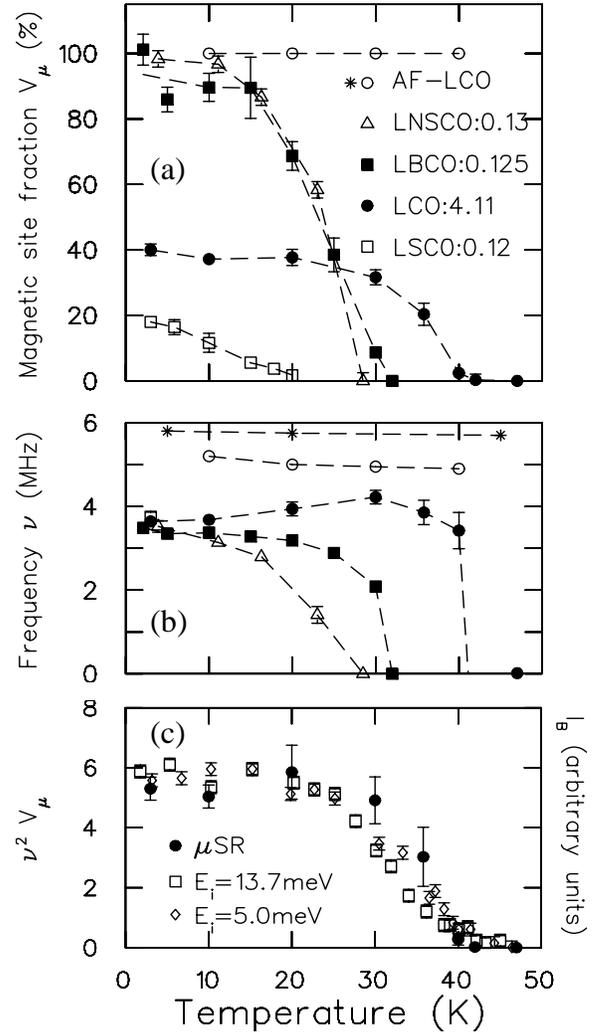,width=3in}}

\vskip 0.2 truein

\label{Figure 7.} 

\caption{(a) Volume fraction $V_{\mu}$ of muon sites
with a static magnetic field larger than $\sim 30$ G
and (b) frequency of the precessing signal in La$_{2}$CuO$_{4.11}$ (LCO:4.11)
and La$_{1.88}$Sr$_{0.12}$CuO$_{4}$ (LSCO:0.12) ([10]),
compared with the results in 
La$_{1.875}$Ba$_{0.125}$CuO$_{4}$ (LBCO:0.125) [28],
La$_{1.47}$Nd$_{0.4}$Sr$_{0.13}$CuO$_{4}$ (LNSCO:0.13) [33], and
antiferromagnetic La$_{2}$CuO$_{4+\delta}$ (AF-LCO) [26].
The broken lines are guides to the eye.
(c) Comparison of the neutron Bragg peak intensity 
$I_{B}$ in La$_{2}$CuO$_{4.11}$ (LCO:4.11)
[31] with those expected from the $\mu$SR results (present study)
as $I_{B} \propto V_{\mu} \times \nu^{2}$.  $\mu$SR and neutron 
results are scaled using the values near $T\rightarrow 0$.  
}

\end{center}

\end{figure}

\newpage

\begin{figure}

\begin{center}

\mbox{\psfig{figure=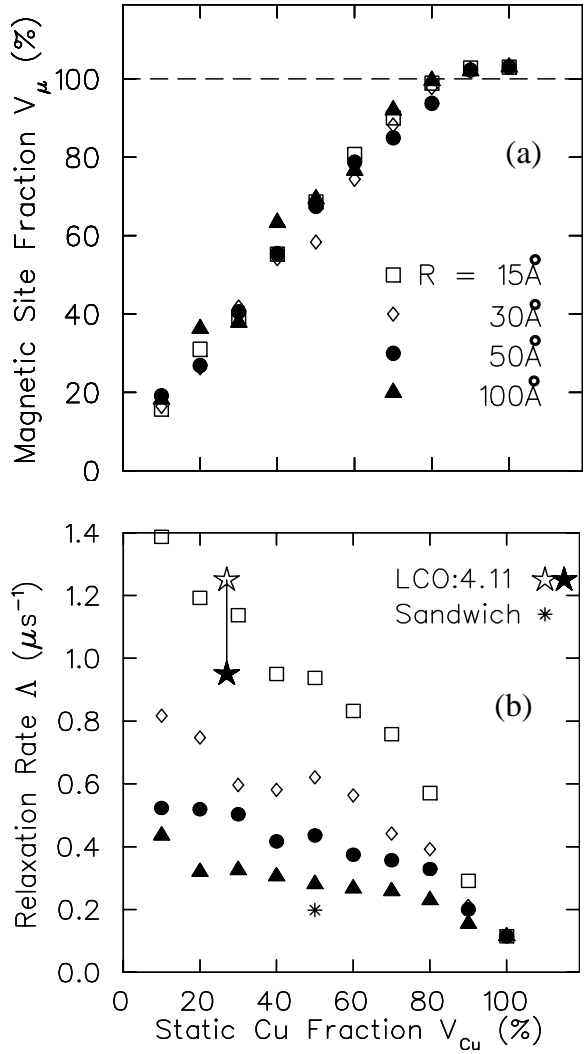,width=3in}}

\vskip 0.2 truein

\label{Figure 8.} 

\caption{(a) Computer simulation results for the volume fraction $V_{\mu}$ 
of muon sites
with a static magnetic field larger than $\sim 30$ G as a function of
the volume fraction of the sample containing static Cu moments, $V_{Cu}$.
(b) The relaxation rate $\Lambda$ of the Bessel function 
oscillation, obtained by fitting the simulation results with Eq. 3, 
plotted as a function of $V_{Cu}$.  Comparison 
with the experimental results for LCO:4.11 (Raw data
$\Lambda_{4.11}$ shown by an open star symbol and corrected data
$\Lambda_{4.11}^{c}$ by a closed star symbol) allows 
estimation of the size of magnetic islands to be
about $R \sim 15-30$ \AA.   The asterisk symbol *
shows the relaxation rate $\Lambda$ expected for the sandwich model.
See ref. [10] fore details. 
}

\end{center}

\end{figure}

\newpage

\begin{figure}

\begin{center}

\mbox{\psfig{figure=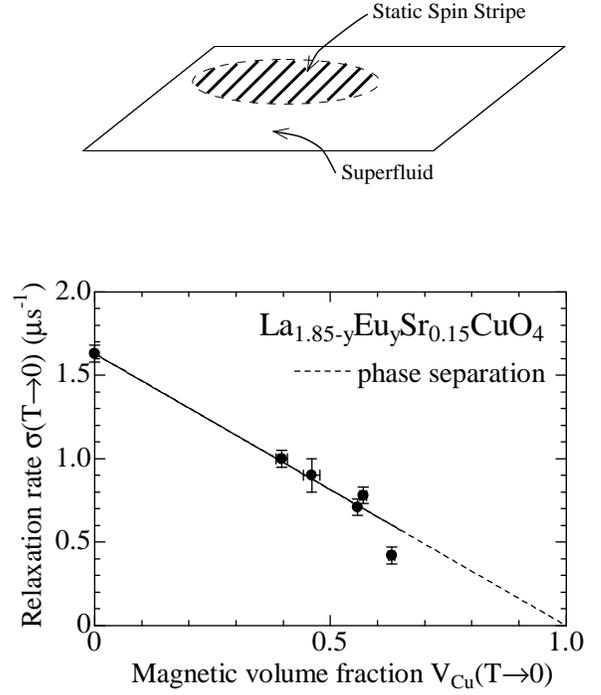,width=3in}}

\vskip 0.2 truein

\label{Figure 9.} 

\caption{Muon spin relaxation rate $\sigma(T\rightarrow 0)$ due to superconductivity
observed in transverse-field $\mu$SR measurements using ceramic specimens 
of La$_{1.85-y}$Eu$_{y}$Sr$_{0.15}$CuO$_{4}$ (LESCO) by
Kojima {\it et al.\/} [11], 
plotted against the volume fraction of static Cu moment
$V_{Cu}$ at $T\rightarrow 0$ determined by zero-field $\mu$SR.  
The trade-off of these two parameters clearly
demonstrates that superconductivity and static magnetism
occur in mutually exclusive volume, as illustrated in the top
figure.}

\end{center}

\end{figure}

\newpage

\begin{figure}

\begin{center}

\mbox{\psfig{figure=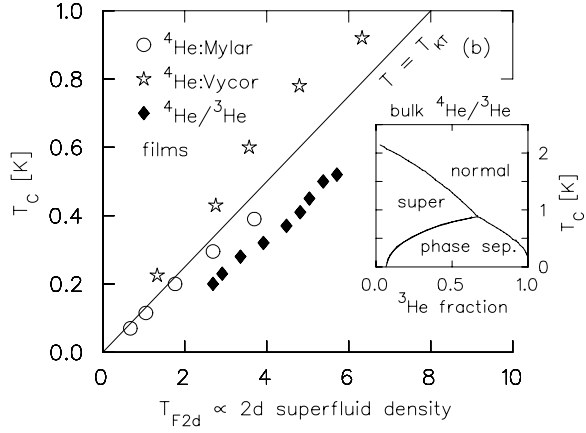,angle=90,width=3in}}

\vskip 0.2 truein

\label{Figure 10.} 

\caption{Superfluid transition temperature $T_{c}$ of $^{4}$He 
films adsorbed on regular (Mylar) [36,37] and porous (Vycor glass) [38] media
and $^{4}$He/$^{3}$He films on fine Alumina powders [39],
plotted against 2-dimensional superfluid density at $T\rightarrow 0$,
after converted to the corresponding 2-dimensional Fermi
temperature [23,35]. Inset figure shows the phase
diagram of bulk $^{4}$He/$^{3}$He mixture.}

\end{center}

\end{figure}

\newpage

\begin{figure}

\begin{center}

\mbox{\psfig{figure=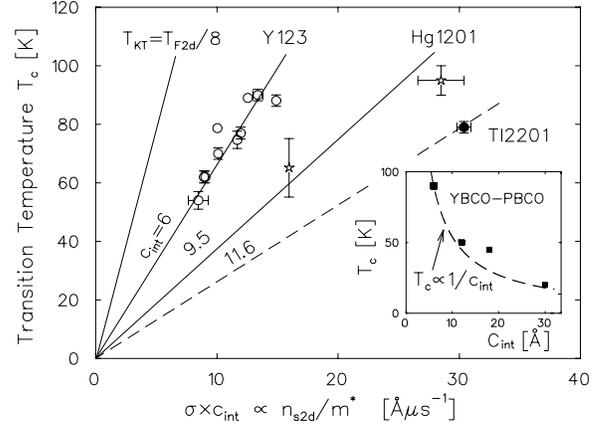,width=3in}}

\vskip 0.2 truein

\label{Figure 11.} 

\caption{Transition temperature $T_{c}$ of HTSC cuprate 
superconductors plotted against the 2-dimensional
superfluid density $n_{s2d}/m^{*}$ obtained
by multiplying the $\mu$SR relaxation rate $\sigma$ with
the average interplaner distance $c_{int}$ between
adjacent CuO$_{2}$ planes [41].  The inset figure
shows variation of $T_{c}$ with $c_{int}$ in 
multi-layer films of YBCO/PrBCO stacked long
the c-axis direction [43]}

\end{center}

\end{figure}

\newpage

\begin{figure}

\begin{center}

\mbox{\psfig{figure=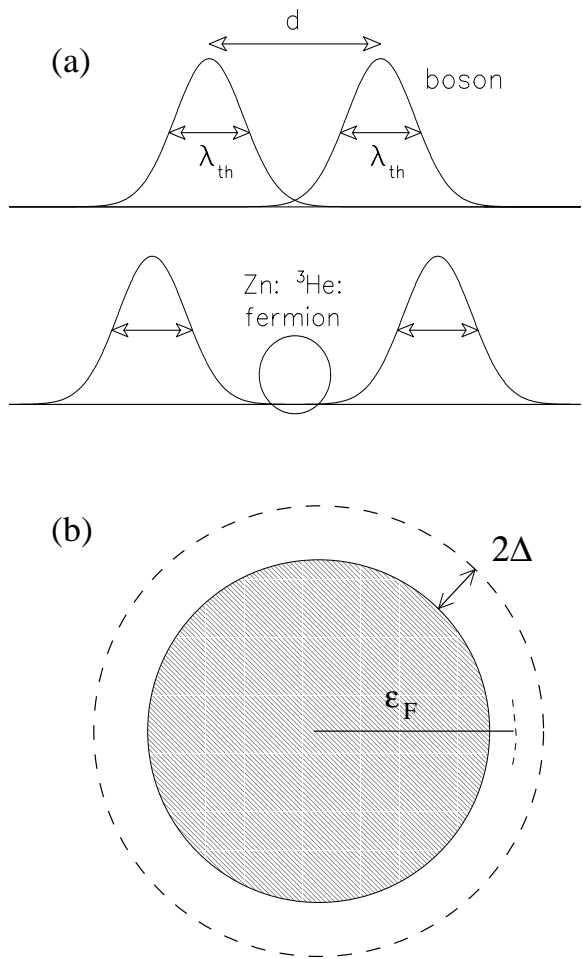,width=3in}}

\vskip 0.2 truein

\label{Figure 12.} 

\caption{(a) A schematic illustration of Bose-Einstein condensation,
which occurs when the wave functions of neighbouring bosons,
with the spread given by the thermal 
de-Broglie wave length $\lambda_{th}$, start to overlap.
In this case $T_{c}$ is directly related to the number density
of bosons.
 (b) Fermi sphere and the energy gap in BCS superconductors.
If one doubles the Debye frequency, the energy gap
$\Delta$
and $T_{c}$ would be doubled, while the superfluid density
(all the carriers in the Fermi sphere) would be unchanged.}

\end{center}

\end{figure}

\newpage

\ \ \  

\newpage

\begin{figure}

\begin{center}

\mbox{\psfig{figure=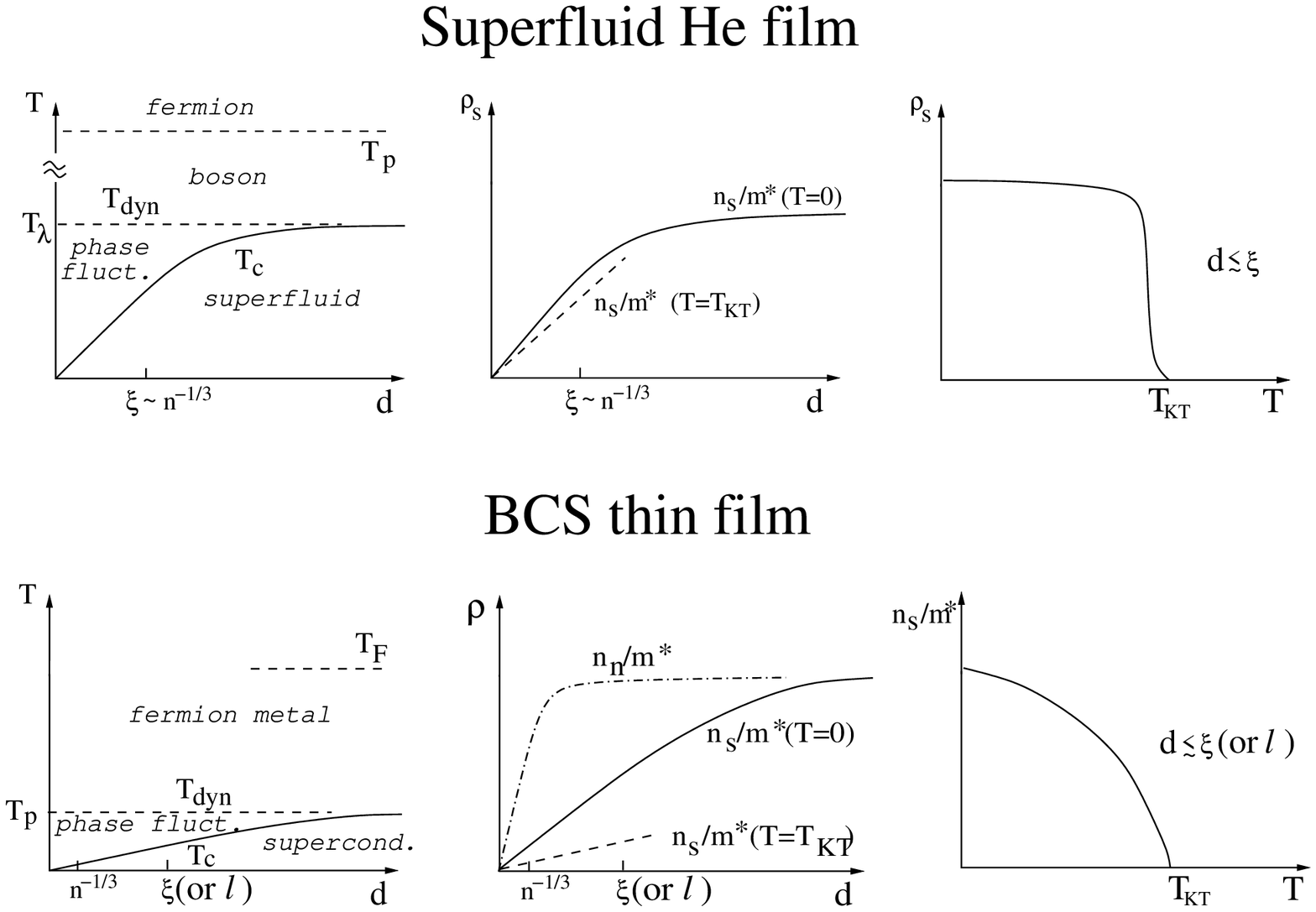,width=6in}}

\vskip 0.2 truein

\label{Figure 13.} 

\caption{Variation of pair-formation temperature scale $T_{p}$, 
the onset temperature $T_{dyn}$ of superfluidity / superconductivity,
the transition temperature $T_{c}$, and the superfluid density,
as a function of film thickness $d$ expected in (a) superfluid He film
and (b) a thin film BCS superconductor.  Also shown
are temperature dependences of the superfluid density
in the 2-d limit where the film thickness is comparable to or less than the
coherence length}

\end{center}

\end{figure}

\newpage

\ \ \ 

\newpage

\begin{figure}

\begin{center}

\mbox{\psfig{figure=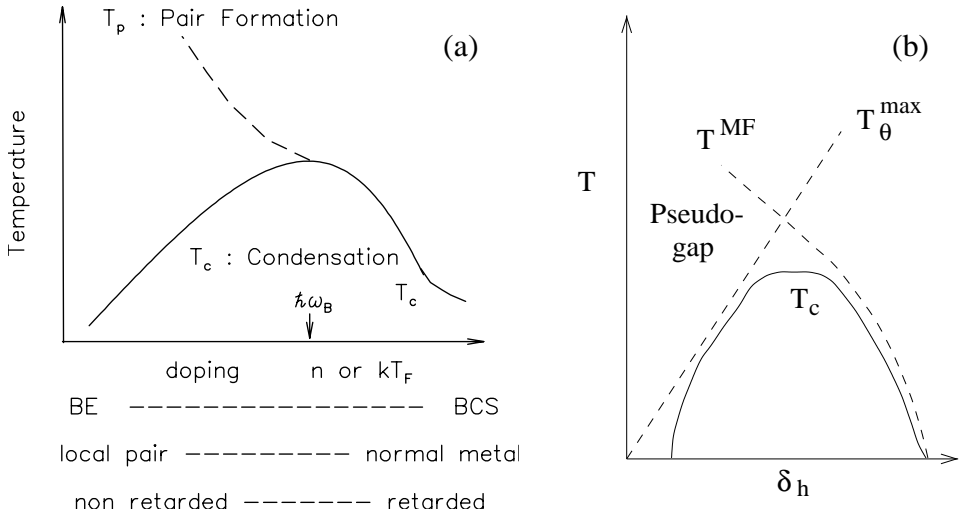,angle=90,width=5in}}

\vskip 0.2 truein

\label{Figure 14.} 

\caption
{Phase diagrams describing (a)
the Bose-Einstein to BCS crossover picture proposed for
HTSC systems by Uemura [47,48,42] and (b) the phase
fluctuation picture proposed by Emery and Kivelson [51]}

\end{center}

\end{figure}

\newpage

\ \ \ 

\newpage

\begin{figure}

\begin{center}

\mbox{\psfig{figure=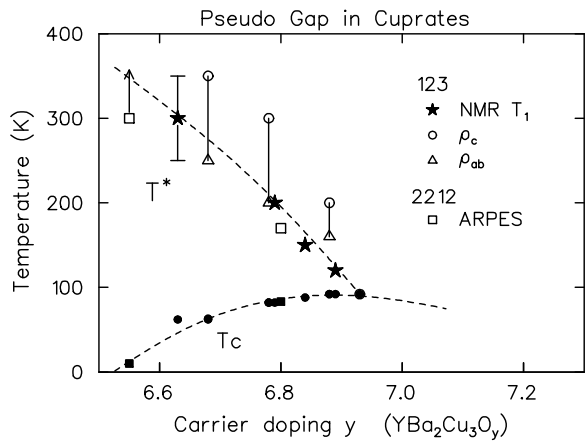,width=3in}}

\vskip 0.2 truein

\label{Figure 15.} 

\caption
{Variation of the pseudogap temperature $T^{*}$ and
the superconducting transition temperature $T_{c}$,
as a function of carrier doping, observed by several
different experimental methods in the Y123 and Bi2212
systems.}

\end{center}

\end{figure}

\newpage

\begin{figure}

\begin{center}

\mbox{\psfig{figure=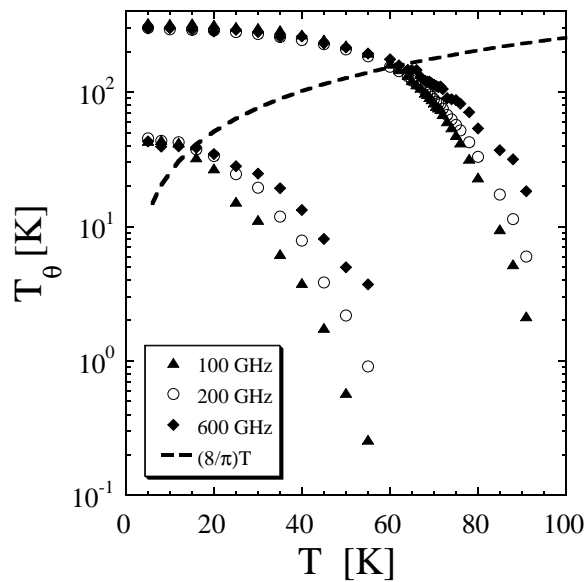,width=3in}}

\vskip 0.2 truein

\label{Figure 16.} 

\caption
{Temperature dependence of the superfluid density
observed in underdoped Bi2212 systems by high-frequency
ac conductivity measurements of Corson {\it et al.\/}
[53] with several different
measuring frequencies.  The broken line shows the
universal relationship between the superfluid density at
$T_{KT}$ and the Kosterlitz-Thouless transition 
temperature $T_{KT}$.}

\end{center}

\end{figure}

\newpage

\begin{figure}

\begin{center}

\mbox{\psfig{figure=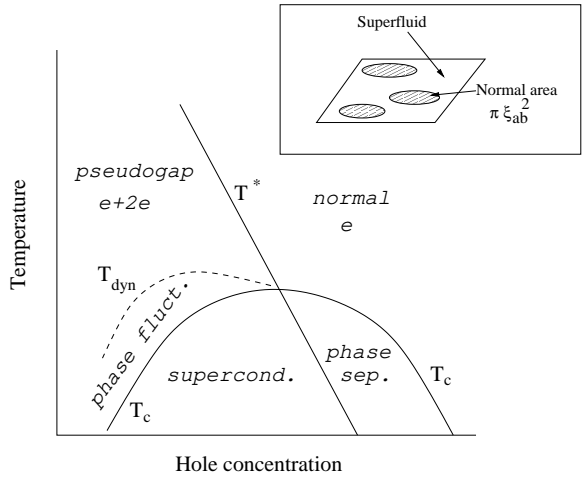,width=3in}}

\vskip 0.2 truein

\label{Figure 17.} 

\caption
{A phase diagram for HTSC cuprates which involves
(1) vanishing of the $T^{*}$ line at a 
critical hole concentration;
(2) $T^{*}$ representing energy scale for 
pair formation;
(3) existence of onset temperature $T_{dyn}$
for dynamic superconductivity with phase fluctuations
(dotted line), and 
(4) microscopic phase separation in the overdoped
region, as illustrated in the inset figure.}

\end{center}

\end{figure}

 \newpage

\begin{figure}

\begin{center}

\mbox{\psfig{figure=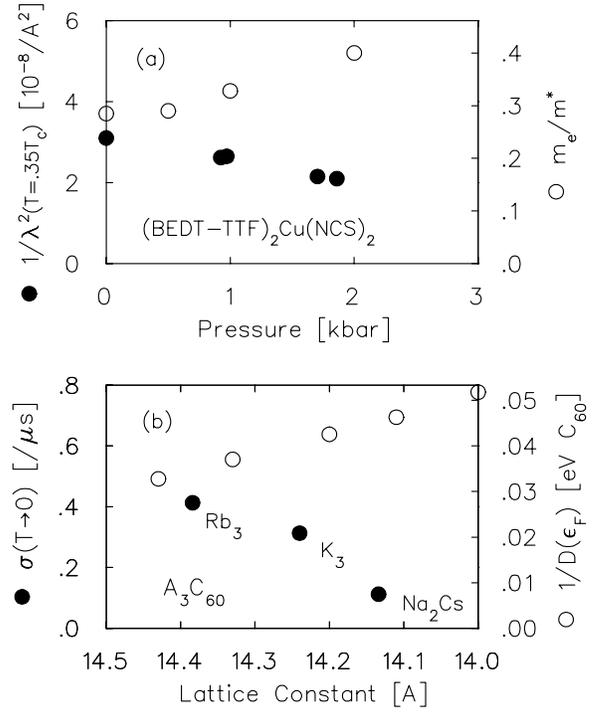,width=3in}}

\vskip 0.2 truein

\label{Figure 18.} 

\caption
{Variation of the superfluid density (closed circle),
obtained from the penetration depth as $1/\lambda^{2}$, 
and the corresponding normal state spectral 
weight $n_{n}/m^{*}$ (open circles) in (a) organic
2-d superconductors (BEDT-TTF)$_{2}$Cu(NCS)$_{2}$
under applied pressure [62,63], and in (b) 
the alkali doped fulleride A$_{3}$C$_{60}$ 
superconductors with varying size of alkali atoms A
[17,18]. The behavior of $n_{n}/m^{*}$ is inferred
from that of $1/m^{*}$ in quantum oscillation 
studies [63] in (a), while the inverse of the density
of states at the Fermi level [63] in A$_{3}$C$_{60}$ }
 
\end{center}

\end{figure}

\end{document}